\documentclass[10pt]{iopart}

\usepackage{bbm}
\usepackage{graphicx}
\usepackage{hyperref}
\usepackage{booktabs}
\usepackage{multirow}

\begin{document}

\title[Optimising the exchange of Majorana zero modes]{Optimising the exchange of Majorana zero modes in a quantum nanowire network}

\author{Tomasz Maciazek}

\address{School of Mathematics, University of Bristol, Fry Building, Woodland Road, Bristol, BS8 1UG, United Kingdom}
\ead{tomasz.maciazek@bristol.ac.uk}

\author{Mia Conlon}
\address{Department of Theoretical Physics, Maynooth University, Ireland}

\begin{abstract}
Determination of optimal control protocols for Majorana zero modes during their exchange is a crucial step towards the realisation of the topological quantum computer. In this paper, we study the finite-time exchange process of Majorana zero modes on a network formed by coupled $p$-wave superconducting one-dimensional nanowires. We provide scalable computational tools for optimising such an exchange process relying on deep learning techniques. To accomplish the scalability, we derive and implement an analytic formula for the gradient of the quantum infidelity which measures the error in the topological quantum gate generation in the Majorana zero modes exchange. Our optimisation strategy relies on learning the optimised transport protocol via a neural net which is followed by direct gradient descent fine tuning. The optimised exchange protocols in the super-adiabatic regime discover the fact that the Majorana zero modes must necessarily stop before crossing a junction point in the network. We explain that this is caused by fast changes in the energy gap of the system whenever one of the Majorana zero modes approaches a junction point. In particular, the energy gap exhibits oscillations followed by a sharp jump. We explain this phenomenon analytically in the regime where the Majorana zero modes are completely localised. Finally, we study how the disorder in the quantum nanowire affects the exchange protocols. This shows that understanding the disorder pattern would allow one to improve quantum gate fidelity by one to two orders of magnitude.
\end{abstract}

%
%
%
%
%

\section{Introduction}
A topological quantum computer realises quantum gates by physically exchanging quantum quasiparticles called anyons \cite{TQC_review}. The possibility of topological quantum computation is a very attractive prospect, because such a computer would be intrinsically robust against local noise \cite{topo_book}. This is because topological quantum gates do not change when the trajectories of the exchanged anyons are perturbed (mathematically, the quantum gate depends only on the homotopy class of the braid). What is more, in systems supporting anyons the quantum computations are realised within a subspace which is energetically gapped, thus any quantum state used in topological quantum computing is characterised by long decoherence times.

One of the main candidates for physical systems able to realise topological quantum computation are systems that support a particular type of anyons called Majorana zero modes (MZMs). There is strong theoretical evidence for the existence of MZMs in two-dimensional $p$-wave superfluids/superconductors \cite{PhysRevLett.98.010506,PhysRevLett.99.037001,PhysRevB.76.104516} assisted by major experimental efforts to realise MZMs in iron-based superconductors \cite{doi:10.1126/science.aan4596,doi:10.1126/science.aao1797,PhysRevB.92.115119,PhysRevB.93.115129,PhysRevLett.117.047001,Kong2021}. However, a practical implementation of topological quantum computation based on MZM braiding has proved to be a major challenge. It is believed that braiding might be accomplished more easily in one-dimensional architectures. In particular, MZMs can also be realised in semiconductor nanowires coupled to a superconductor \cite{PhysRevLett.105.077001,PhysRevLett.105.177002,Alicea2011,Das2012,doi:10.1126/science.1222360} as well as in other condensed matter and photonics systems \cite{doi:10.1126/science.1259327,Xu2016} which provide experimental realisations of Kitaev's one-dimensional superconductor model \cite{Kitaev_chain}. There, the MZMs are localised at the endpoints of the topological regions in the nanowire and can be transported along the wire adiabatically by tuning local voltage gates distributed along the wire \cite{Alicea2011}. If several such nanowires are coupled together to form a junction (or more generally, a network), then MZM braiding can be accomplished by adiabatic transport through the junction. Remarkably, such a braiding of MZMs moving on 1D nanowire networks  produces the same quantum gates as MZM braiding in 2D \cite{Alicea2011,PhysRevB.84.035120}. Although the quantum gates that can be obtained by MZM braiding do not permit universal quantum computation \cite{TQC_review,PhysRevA.71.022316}, their implementation would constitute a critical step towards the realisation of a universal topological quantum computer.

Our presented work concerns the general issue of optimal control of MZMs in networks 1D quantum nanowires. Similar topics have been studied before in relation to the quantum control of MZMs in a single wire \cite{PhysRevB.88.064515,PhysRevB.91.201404,PhysRevB.100.134307,LuukNN}. The general objective is to transport the MZMs in a given finite amount of time so that the MZM motion profiles (the MZM positions in time) maximise the quantum fidelity. Even though the MZMs are protected by the energy gap, such finite time manipulations may cause leaking of the quantum state to the higher energy levels. Although exponentially small for slow manipulations, these effects are important as they may constitute the ultimate source of errors. In order to mitigate such non-adiabatic transitions in different non-adiabatic motion regimes, the {\textit{bang-bang}} \cite{PhysRevB.91.201404} and {\textit{jump-move-jump}} protocols \cite{LuukNN} have been proposed. In this paper, we work exclusively in the (super)-adiabatic regime where the evolution times are much longer than the inverse of the energy gap and the MZM velocity is smaller than the superconducting order parameter \cite{PhysRevB.100.134307}. In this regime, when the MZMs are transported by small distances in a single wire, the optimal transport protocol is the simple {\textit{ramp-up/ramp-down}} protocol \cite{PhysRevB.88.064515}. Although MZM transport in a single wire is well-studied, it turns out that optimising the full exchange process poses its own challenges and has its distinct features. Firstly, there is the technical difficulty of efficiently simulating long time quantum evolution and computing the gradient of the quantum fidelity for such a long-time process. To overcome this, we build a scalable machine learning system (a neural net with three hidden layers, inspired by the approach of the work \cite{LuukNN}) where the gradient of the quantum fidelity with respect to MZM positions in time is computed analytically. This allows us to mitigate the so-called caching problem in automatic differentiation, thus significantly reducing the required amount of RAM (which would otherwise be a significant bottleneck issue, effectively allowing for simulations of only extremely small systems). Our code is openly available online \cite{github_repo}. Secondly, as we explain in Sections \ref{sec:learning} and \ref{sec:gap}, the energy gap exhibits complex behaviour when one of the MZMs crosses a junction point during the exchange. In particular, the energy gap oscillates and jumps sharply  (although the amplitude of these oscillations is much smaller than the amplitude of the energy gap jump). This means that even in case when the MZMs move adiabatically when being located far away from the junction point, the time derivative of system's Hamiltonian may become large when one of the MZMs approaches the junction, making the adiabatic evolution difficult to maintain. Due to this effect, optimising the entire exchange protocol is a nontrivial task. As we show in Section \ref{sec:learning}, the optimised exchange profiles share one common feature, namely they require the MZMs to slow down and stop before crossing the junction.

Machine learning (ML) techniques have recently seen a surge of applications to condensed matter physics, see e.g. \cite{Schmidt2019,Bedolla_2021,dawid2022modern} for recent reviews. Of particular relevance in the context of our presented work is \cite{LuukNN}, where they use ML techniques to study the optimisation of shuttling a MZM along a wire. In addition in  \cite{PhysRevLett.130.116202} and \cite{dassarma_RNN}, ML techniques have been used to optimise the design of nanowire-based systems supporting MZMs and to predict the profile of disorder in a nanowire. We return to these topics in Section 4. What is more, reinforcement learning is applied to optimise the compilation of an arbitrary qubit gate into a sequence of elementary braiding moves \cite{PhysRevLett.125.170501}.

This paper is structured as follows. In Section \ref{sec:theory} we describe the theoretical setup for studying quantum control of MZMs. In Section \ref{sec:fidelity} we take a closer look at quantum fidelity and compute its gradient analytically. In Section \ref{sec:learning} we present details of the numerical optimisation protocol, present the optimised exchange profiles and discuss the effects of disorder in the nanowire. In Section \ref{sec:gap} we explain the aforementioned jump in the energy gap and discuss its consequences for the shape of the optimised exchange profiles.

\section{Theoretical setup}\label{sec:theory}
\begin{figure}
    \centering
    \includegraphics[width=0.55\textwidth]{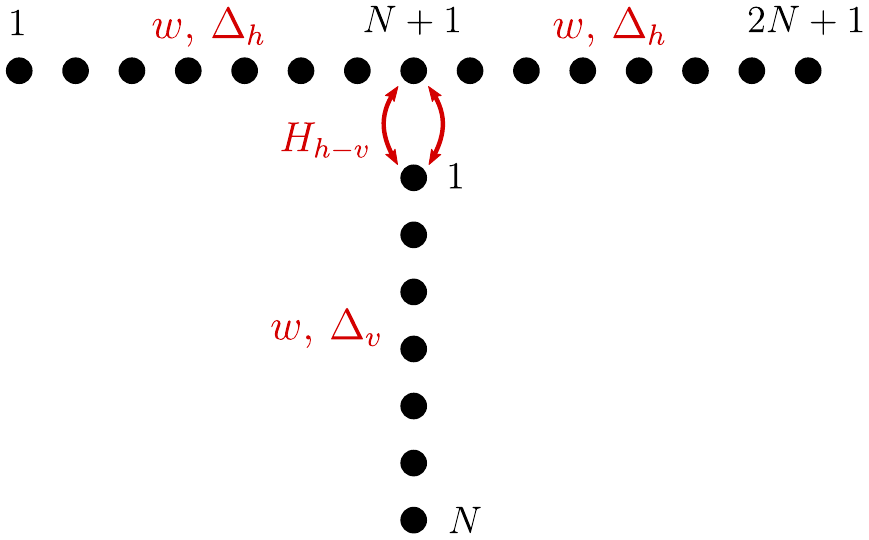}
    \caption{The trijunction setup. We assume that the horizontal and vertical chains have identical hopping amplitudes, but possibly different superconducting order parameters, $\Delta_h$ and $\Delta_v$ respectively. Note the site labelling convention where the site $N+1$ of the horizontal chain couples to the site $1$ of the vertical chain.}
    \label{fig:trijunction}
\end{figure}

A trijunction consists of two chains (see also \Fref{fig:trijunction}): i) the horizontal chain of the length $2N+1$ whose Hamiltonian reads
\begin{equation}\label{eq:ham_h}
\eqalign{
& H_h(t) = -\sum_{j=1}^{2N+1} \mu_j^{(h)}(t) \left(c_{j}^\dagger c_j - \frac{1}{2}\right) - w \sum_{j=1}^{2N} \left(c_{j}^\dagger c_{j+1} + c_{j+1}^\dagger c_{j}\right) \\ 
& + \sum_{j=1}^{2N} \left(\Delta_h\, c_{j} c_{j+1} + \overline \Delta_h\, c_{j+1}^\dagger c_{j}^\dagger\right)
}
\end{equation}
and ii) the vertical chain of the length $N$ whose Hamiltonian reads
\begin{equation}\label{eq:ham_v}
\eqalign{ 
& H_v(t) = -\sum_{j=1}^{N} \mu_j^{(v)}(t) \left(d_{j}^\dagger d_j - \frac{1}{2}\right) - w \sum_{j=1}^{N-1} \left(d_{j}^\dagger d_{j+1} + d_{j+1}^\dagger d_{j}\right) \\ 
& + \sum_{j=1}^{N-1} \left(\Delta_v\, d_{j} d_{j+1} + \overline \Delta_v\, d_{j+1}^\dagger d_{j}^\dagger\right),
}
\end{equation}
where the on-site potentials have the forms
\begin{equation}
\mu_j^{(h)}(t) = \mu_0 - V_j^{(h)}(t), \quad \mu_j^{(v)}(t) = \mu_0 - V_j^{(v)}(t).
\end{equation}
The Hamiltonian of the entire system is given by
\begin{equation}\label{eq:Htotal}
H(t) = H_h(t) + H_v(t) + H_{h-v},
\end{equation}
where $H_{h-v}$ is the coupling between the site $N+1$ of the horizontal chain and the site $1$ of the vertical chain. We consider the coupling of the form
\begin{equation}\label{eq:coupling}
\fl H_{h-v} = -w\left(c_{N+1}^\dagger d_{1} + d_{1}^\dagger c_{N+1}\right)+  \left(\Delta_v\, c_{N+1} d_{1}+ \overline \Delta_v\, d_{1}^\dagger c_{N+1}^\dagger\right).
\end{equation}
Recall that, depending on the relationships between the coefficients $\{\mu_j^{(h/v)}\}$, $\Delta_{h/v}$, and $w$, different regions of the trijunction may be in different phases. In particular, in the region where $|\mu^{(h/v)}_j|<2w$ and $\Delta_{h/v}\neq 0$ the system is in the topological phase with MZMs localised on the boundary of this region \cite{Kitaev_chain}. On the other hand, if $|\mu^{(h/v)}_j|>2w$, then no MZMs appear in the corresponding region and the system is in the topologically trivial phase. In the numerical calculations in Sections \ref{sec:learning} and \ref{sec:gap} we take $\Delta_h \equiv \Delta >0$ and $\Delta_v = i\Delta$. Note that we cannot assume the superconducting order parameters to be real numbers everywhere in the system, as it would inevitably cause the existence of a $\pi$-junction during the exchange process causing level crossings and creating extra pairs of MZMs, see e.g. \Sref{sec:gap} or \cite{Alicea2011} for more details.

\begin{figure}
    \centering
    \includegraphics[width=0.5\textwidth]{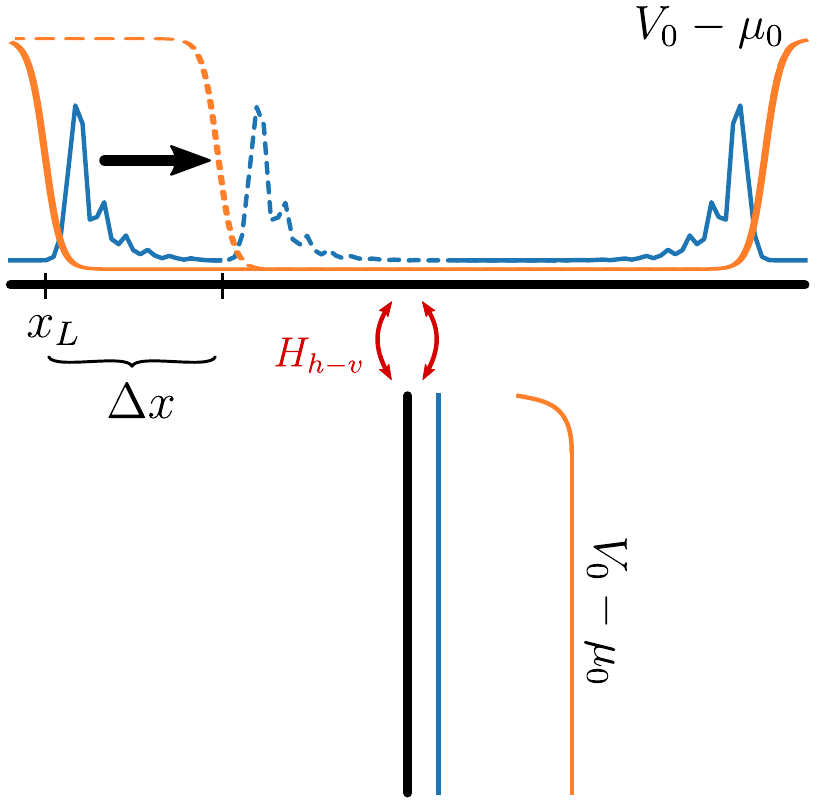}
    \caption{The MZMs are transported by shifting the positions of the sigmoid functions that determine the on-site potentials. In the figure, the on-site potentials are represented by the orange lines and the MZM amplitudes by the blue lines. The plots are largely schematic, but they represent MZMs in the Hamiltonian used in Section \ref{sec:learning}. This particular configuration of MZMs takes place in stages $I$ and $IV$ of the exchange. $H_{h-v}$ represents the coupling across the junction, see \Eref{eq:coupling}.}
    \label{fig:sigmoid}
\end{figure}

By changing the on-site potentials $V_j^{(h)}(t)$ and $V_j^{(v)}(t)$ one can control the MZMs so that they will move around the network. This can be realised experimentally in the keyboard-architecture setup for controlling MZMs \cite{Alicea2011,10.21468/SciPostPhys.5.1.004,PhysRevB.107.104516} which physically corresponds to distributing voltage gates along the wire which are tuned whenever the local voltage needs to be changed. To model this, we assume $V_j^{(h/v)}$ to have the shape of the sigmoid function. 
\begin{figure}
    \centering
    \includegraphics[width=0.75\textwidth]{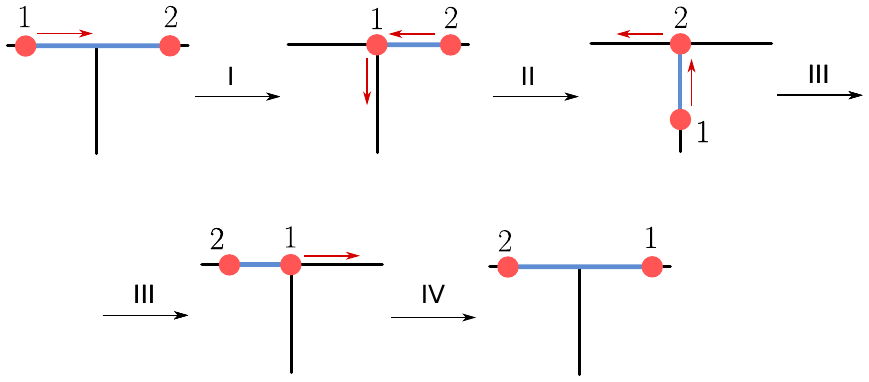}
    \caption{The four stages of the MZM exchange. The blue strings denote the topological regions and the red dots denote the positions of the MZMs. In each stage, the positions of the MZMs are determined by the vectors $\mathbf{s}_j^{(*)}$ of the length $N_T$, where $j=1,2$ labels the MZMs and $* = I, II, III, IV$ labels the stages. The vectors $\mathbf{s}_j^{(*)}$ determine the MZM positions according to Equations \eref{eq:s_to_x_I}- \eref{eq:s_to_x_III} presented in \Sref{sec:learning}.}
    \label{fig:stages}
\end{figure}
For instance, to place one topological region in the horizontal chain (stages $I$ and $IV$ of the exchange, see \Fref{fig:stages}), we set
\begin{equation}\label{eq:potential_sigmoid_I_IV}
\eqalign{
\fl V_j^{(h)} =V_0 \left(\sigma(j - x_R^{(I/IV)}) + \sigma(x_L^{(I/IV)} - j)\right),\quad j=1,\dots,2N+1, \cr
\fl V_j^{(v)} =V_0\,\sigma(j),\quad j=1,\dots, N,
}
\end{equation}
where $0\leq x_L^{(I/IV)}<x_R^{(I/IV)}\leq 2N+1$ are the (approximate) positions of the MZMs, $\mu_0$ is the uniform background potential satisfying $|\mu_0|<2w$ and
\[ \sigma(x) = \frac{1}{1+e^{-x}},\quad V_0 > 2w+\mu_0.\]
 In order to move the left MZM to the right by some distance $\Delta x$ in time $T$, we parametrise $x_L(s) = x_L + s\Delta x$, $s = t/T$, see \Fref{fig:sigmoid}. 
 Similarly, in the configuration where one of the MZMs is on the vertical chain and the other one is on the right side of the horizontal chain (stage $II$ of the exchange) , the on-site potentials read
\begin{equation}\label{eq:potential_sigmoid_II}
\eqalign{
\fl V_j^{(v)} =V_0 \left(\sigma(j - x_R^{(II)}) + \sigma(j-x_V^{(II)}+1)\right),\quad j=1,\dots, N \cr
\fl V_j^{(h)} =V_0 \left(\sigma(j - x_R^{(II)}) + \sigma(N-x_V^{(II)} -j+1)\right),\quad j=N+1,\dots, 2N+1 \cr
\fl V_j^{(h)} =V_0\,\sigma(N+1-j),\quad j=1,\dots, N,
}
\end{equation}
where $N+1\leq x_R^{(II)}\leq 2N+1$ and $1\leq x_V^{(II)}\leq N$. Finally, in the configuration where one of the MZMs is on the vertical chain and the other one is on the left side of the horizontal chain (stage $III$ of the exchange), the on-site potentials read
\begin{equation}\label{eq:potential_sigmoid_III}
\eqalign{
\fl V_j^{(v)} =V_0 \left(\sigma(x_L^{(III)}-j-N-1) + \sigma(j-x_V^{(III)})\right),\quad j=1,\dots, N \cr
\fl V_j^{(h)} =V_0 \left(\sigma(x_L^{(III)}-j) + \sigma(j-x_V^{(III)} -N-1)\right),\quad j=1,\dots, N+1 \cr
\fl V_j^{(h)} =V_0\,\sigma(j-N-1),\quad j=N+2,\dots, 2N+1,
}
\end{equation}
where $1\leq x_L^{(III)}\leq N+1$ and $1\leq x_V^{(III)}\leq N$.

We work exclusively in the (super)-adiabatic regime where the evolution time $T$ is much larger than the inverse of the energy gap of the system and the MZM velocity is smaller than the critical velocity $v_{crit}=\Delta$, i.e.
\begin{equation}\label{eq:adiabatic_conditions}
v<v_{crit} = \Delta,\qquad T> \frac{2\pi}{E_{gap}}.
\end{equation}

In the numerical calculations we make use of the Bogolyubov-de-Gennes form of the Hamiltonian \eref{eq:Htotal}, which is the Hermitian matrix $H_{BdG}$ such that
\begin{equation}\label{eq:H_bdg}
H(t) = \frac{1}{2}
\pmatrix{
\mathbf{C}^\dagger & \mathbf{D}^\dagger & \mathbf{C} & \mathbf{D}
}
\, H_{BdG}(t)\, 
\pmatrix{ 
\mathbf{C} \cr
 \mathbf{D} \cr
\mathbf{C}^\dagger  \cr
  \mathbf{D}^\dagger
 },
\end{equation}
where $\mathbf{C}^T = \left(c_1\dots,c_{2N+1}\right)$ and $\mathbf{D}^T= \left(d_1,\dots,d_{N}\right)$. The Bogolyubov-de-Gennes Hamiltonian can be diagonalised as
\begin{equation}\label{eq:diagonalisation}
W(t)^\dagger H_{BdG}(t) W(t) = E(t),
\end{equation}
where $E(t)$ is the diagonal matrix containing the single-particle spectrum and $W(t)$ is the matrix of eigenmodes, i.e. the Bogolyubov transformation diagonalising $H_{BdG}(t)$. Recall that such a fermionic Bogolyubov transformation has the form \cite{BdGbook}
\begin{equation}
W(t) = 
\pmatrix{
U(t) & \overline V(t) \cr
V(t) & \overline U(t)
},
\end{equation}
where $U(t)$ and $V(t)$ are blocks of the size $(3N+1)\times(3N+1)$.

\section{Quantum fidelity and its gradient}\label{sec:fidelity}
When exchanging the MZMs, we consider the $p$-wave Hamiltonian that changes in time, $H(t)$, $0\leq t\leq T$ such that $H(0)=H(T)$. In order to simulate the quantum evolution, we divide the time interval into $N_T$ timesteps, each timestep having the length $\Delta t = T/N_T$. We approximate the quantum evolution by the Suzuki-Trotter formula in the BdG picture
\begin{equation}\label{eq:Oev}
\mathcal{O}_{ev} \approx \prod_{k=1}^{N_T} \exp\left(-i\Delta t H_{BdG}(t_{N_T-k+1})\right),\quad t_j = j\,\Delta t,
\end{equation}
where we employ the convention that the evolution in the first timestep corresponds to the far-right element of the product. The quantum fidelity compares the evolved eigenmodes $W_{ev} = \mathcal{O}_{ev}W(0)$ with the target eigenmodes $W(T)$ from \Eref{eq:diagonalisation}. In particular, we define the quantum fidelity as the overlap between the Bogolyubov vacuums corresponding to the eigenmodes of $H(t)$ at the evolved eigenmodes $W_{ev}(T)$. The result is given by the Onishi formula \cite{Onishi66,Beck1970,BdGbook}
\begin{equation}\label{eq:onishi}
\mathcal{F} =\left| \det\left[\, U(T)^\dagger U_{ev}  + V(T)^\dagger V_{ev} \right] \right|,
\end{equation}
where $U_{ev}$ and $V_{ev}$ are the respective blocks of $W_{ev}$.

Let us next briefly explain how to compute the gradient of the quantum fidelity. To this end, we use the following two identities  
\begin{equation}
\fl \partial_\lambda\, |z(\lambda)| = \frac{1}{|z|}\, \Re\left(\overline z\, \partial_\lambda z\right), \quad \partial_\lambda\det X(\lambda) = \det X\, \tr\left(X^{-1}\,\partial_\lambda X\right).
\end{equation}
The role of the parameter $\lambda$ in \Eref{eq:onishi} is played by the entries of the vectors $\mathbf{s}_{j}^{(*)}$, $j = 1,2$ and $* = I,II, III, IV$. Note that only the matrices $U_{ev}$ and $V_{ev}$ depend on such a $\lambda$, since they come from the quantum evolution operator. What is more, they depend linearly on $\mathcal{O}_{ev}$ as follows.
\begin{equation}
U_{ev} = P_1 \mathcal{O}_{ev} W(0)P_1^T,\quad V_{ev} = P_2 \mathcal{O}_{ev} W(0) P_1^T,
\end{equation}
where $P_1 = \left(\mathbbm{1}, 0\right)$, $P_2 = \left(0, \mathbbm{1}\right)$ with $\mathbbm{1}$ and $0$ being matrices of the sizes $(3N+1)\times (3N+1)$. Thus, we can express the gradient of the quantum fidelity in terms of the gradient of the quantum evolution operator as
\begin{equation}
\fl \eqalign{\partial_\lambda\mathcal{F} = \mathcal{F}\, \Re\Big{\{}\tr\left[\left(U(T)^\dagger U_{ev}  + V(T)^\dagger V_{ev}\right)^{-1}\right] \left(U(T)^\dagger P_1  + V(T)^\dagger  P_2\right) \times \cr
\times \left(\partial_\lambda\mathcal{O}_{ev}\right) W(0) P_1^T \Big{\}}.
}
\end{equation}
As we mentioned earlier in this section, we are interested in computing the gradient of $\mathcal{F}$ with respect to the positions of MZMs in each timestep. If $\lambda$ is the position of MZM with label $1$ in the $k$-th timestep, i.e. $\lambda = s_{1,i}^{(*)}$ for some $i$ and $* = I,II, III, IV$ (here, to simplify the notation, $k$ enumerates all the timesteps collectively, while $i$ enumerates only the timesteps within a given exchange stage), then the derivative $\partial_\lambda$ affects only the $k$-th term in the product \eref{eq:Oev}, i.e.
\begin{equation}\label{eq:D_Oev}
\partial_\lambda\mathcal{O}_{ev} = \mathcal{O}_{ev}^{(k,+)} \left(\partial_\lambda e^{-i H_{BdG}(t_k,\lambda)\Delta t}\right)\mathcal{O}_{ev}^{(k,-)},
\end{equation}
where $\mathcal{O}_{ev}^{(k,-/+)} $ are the evolution operators before and after the timestep $k$
\[
\fl \mathcal{O}_{ev}^{(k,-)} = \prod_{l=1}^{k-1} \exp\left(-i\Delta t H_{BdG}(t_{k-l})\right),\quad \mathcal{O}_{ev}^{(k,+)} = \prod_{l=k+1}^{N_T}  \exp\left(-i\Delta t H_{BdG}(t_{N_T-l+k+1})\right).
\]
Finally, the task at hand boils down to computing the derivative of the exponent in the \Eref{eq:D_Oev}. This is a standard problem and there exist several techniques to address it. In this work, we choose to apply the following method \cite{JB76,KL85,TC03}.
\begin{equation}
\partial_\lambda e^{-i H_{BdG}(t_k,\lambda)\Delta t} = W(t_k) X_k W(t_k)^\dagger,
\end{equation}
where the $(p,q)$-th entry of $X_k$ reads
\begin{equation}
\eqalign{
i g_{p,q}^{(k)}\, \frac{e^{-i\Delta t\, \epsilon(t_k)_{p}}-e^{-i\Delta t\, \epsilon(t_k)_{q}}}{ \epsilon(t_k)_{p}- \epsilon(t_k)_{p}},\quad p\neq q \cr
g_{p,p}^{(k)}\, \Delta t\, e^{-i\Delta t\, E(t_k)_{p,p}},\quad p = q,
}
\end{equation}
and $\epsilon(t_k)_{p}$ is the $p$-th diagonal entry of $E(t_k)$ and $g_{p,q}^{(k)}$ is the $(p,q)$-th entry of 
\[G^{(k)} = -i\, W(t_k)^\dagger \left(\partial_\lambda H_{BdG}(t_k,\lambda)\right)W(t_k).\]
Finding the derivative $\partial_\lambda H_{BdG}(t_k,\lambda)$ is a straightforward task, because only the diagonal entries of $H_{BdG}(t_k)$ depend on the positions of the MZMs and the dependency has the form of the simple sigmoid function, as shown in Equations \eref{eq:potential_sigmoid_I_IV}, \eref{eq:potential_sigmoid_II} and \eref{eq:potential_sigmoid_III}.

\section{Machine learning the optimised transport profiles}
\label{sec:learning}
\begin{figure}
    \centering
    \includegraphics[width=0.9\textwidth]{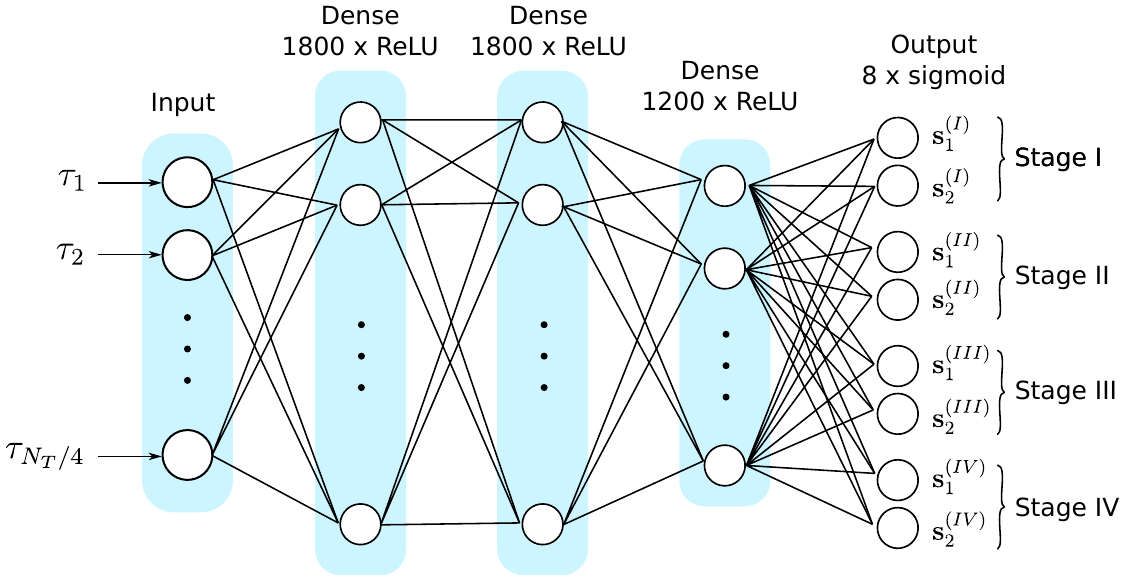}
    \caption{The neural net architecture that we used for optimising the MZM transport profiles. The neural net has three hidden layers of $1800$, $1800$ and $1200$ ReLU units respectively. The eight-unit sigmoid output layer determines the MZM transport profile. The cost function is the quantum infidelity, i.e. $1-\mathcal{F}$. The input of the NN is fixed to be the vector of $N_T/4$ evenly spaced numbers over the interval $[0,1]$. The output vectors determine the positions of the MZMs in the respective stages.}
    \label{fig:nn_schematic}
\end{figure}

The strategy for optimising the transport profiles is twofold. Firstly, we use a neural net (NN) with eight sigmoid output neurons to generate the vectors $\mathbf{s}_j^{(*)}$ of the length $N_T/4$, where $j=1,2$ labels the MZMs and $* = I, II, III, IV$ labels the stages (see \Fref{fig:nn_schematic}). The NN architecture presented in \Fref{fig:nn_schematic} has been determined as suitable for the problem at hand by trial and error iterations over different NN depths and hidden layer widths.
The input of the NN, $\tau$, is fixed to be the vector of $N_T/4$ evenly spaced numbers over the interval $[0,1]$. The output vectors determine the positions of the MZMs in the respective stages according to Equations \eref{eq:s_to_x_I}-\eref{eq:s_to_x_III} below. The cost function for the neural net  training is the quantum infidelity, $1-\mathcal{F}$, (see \Eref{eq:onishi}) whose gradient with respect to NN's outputs we calculate analytically and subsequently backpropagate through the NN. We use the Adam optimiser \cite{Adam} with the learning rate $10^{-4}$. Secondly, after training the NN for $100 - 200$ episodes, we fine tune the resulting profiles by running the gradient descent directly in the space of the vectors $\mathbf{s}_j^{(*)}$ with the Adam optimiser and the learning rate $10^{-6}$. We have found such a procedure to be most effective, because the NN is able to efficiently optimise the global shapes of the transport profiles with different layers of the NN learning the features of the curves on different scales. The smaller-scale fine tuning is most effectively done using the direct gradient descent. 

Let us next specify how the positions of MZMs in Equations \eref{eq:potential_sigmoid_I_IV}, \eref{eq:potential_sigmoid_II} and \eref{eq:potential_sigmoid_III} are determined by the NN output vectors $\mathbf{s}_j^{(*)}$ (recall also our convention for labelling the sites of the chains in \Fref{fig:trijunction}). The exchange stages are schematically shown in \Fref{fig:stages}. In stages $I$ and $IV$ of the exchange both MZMs are located on the horizontal chain, and the potential profile is given by \Eref{eq:potential_sigmoid_I_IV}. The positions of the MZMs for each time step in stage $I$ read (note that we use vector notation where $\mathbf{x}_{L/R}^{(*)}$ are vectors of the length $N_T/4$ that contain the positions of the MZMs in each time step)
\begin{equation}\label{eq:s_to_x_I}
\eqalign{
 \mathbf{x}_L^{(I)} = x_0 + \mathbf{s}_1^{(I)}(N+1-x_0), \cr
 \mathbf{x}_R^{(I)} = \mathbf{s}_2^{(I)}(N+1) + \left(1-\mathbf{s}_2^{(I)}\right)(2N+1-x_0),
}
\end{equation}
where $x_0$ is the smallest distance at which the MZMs are allowed to near the edge of the system. For the calculations presented in this section we took $x_0=5$. In stage $IV$ the positions of the MZMs are switched, so we have
\begin{equation}\label{eq:s_to_x_IV}
\eqalign{
 \mathbf{x}_L^{(IV)} = \mathbf{s}_2^{(IV)} x_0 + \left(1-\mathbf{s}_2^{(IV)}\right)(N+1), \cr
 \mathbf{x}_R^{(IV)} = \left(1-\mathbf{s}_1^{(IV)}\right)(N+1) + \mathbf{s}_1^{(IV)}(2N+1-x_0).
}
\end{equation}
In stage $II$ the MZM with label $1$ is located on the vertical chain while the MZM with label $2$ is located on the right half of the horizontal chain. The potential profile in this stage has the form \eref{eq:potential_sigmoid_II}, where
\begin{equation}\label{eq:s_to_x_II}
\eqalign{
 \mathbf{x}_V^{(II)} = 1-\mathbf{s}_1^{(II)} + \mathbf{s}_1^{(II)} (N-x_0), \cr
 \mathbf{x}_R^{(II)} = \mathbf{s}_2^{(II)}(N+1) + \left(1-\mathbf{s}_2^{(II)}\right)(2N+1-x_0).
}
\end{equation}
In stage $III$ the MZM with label $1$ is located on the vertical chain while the MZM with label $2$ is located on the left half of the horizontal chain. The potential profile in this stage has the form \eref{eq:potential_sigmoid_III}, where
\begin{equation}\label{eq:s_to_x_III}
\eqalign{
 \mathbf{x}_V^{(III)} = \left(1-\mathbf{s}_1^{(III)}\right)(N-x_0) + \mathbf{s}_1^{(III)}, \cr
 \mathbf{x}_L^{(III)} = \left(1-\mathbf{s}_2^{(III)}\right)(N+1) + \mathbf{s}_2^{(III)}x_0.
}
\end{equation}

\begin{figure}
    \centering
    \includegraphics[width=\textwidth]{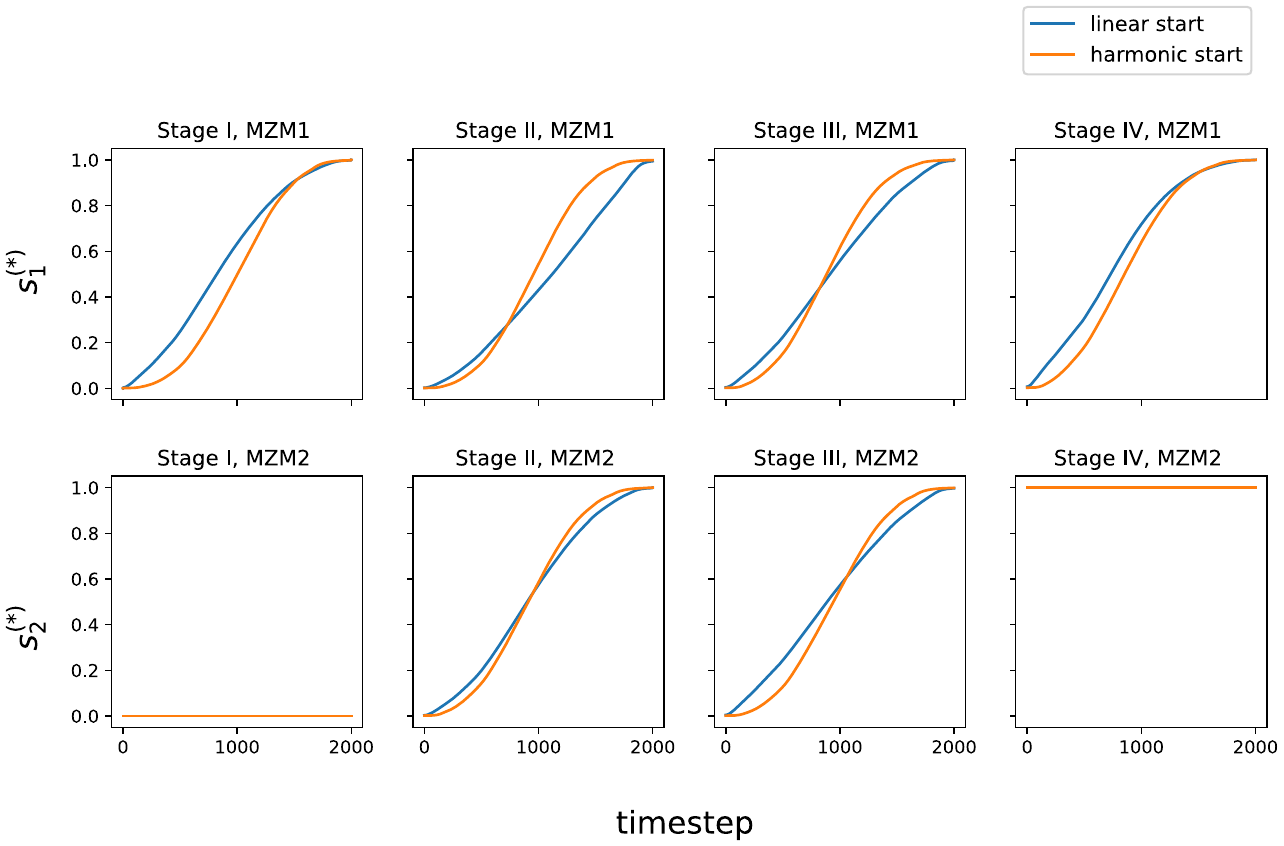}
    \caption{The optimised exchange profiles for the system of the size $3N+1$ with $N=55$, $\Delta=0.55$, $w=2.0$, $\mu_0=1.0$, $V_0 = 30.1$, $x_0=5$. The evolution time of each stage was $250.0$ (total time $T=1000.0$) and consisted of $2000$ time steps (total number of time steps $N_T=8000$). We have considered two different starting profiles: the linear motion (no stops at the junction) and the harmonic motion (with stops at the junction). The optimisation consisted of $120$ epochs of NN training which reduced the infidelities from $0.3668$ to $0.0017$ for the linear motion start and from $0.0744$ to $0.0008$ for the harmonic motion start. The NN training was followed by $60$ gradient descent steps directly in the space of $\mathbf{s}_j^{(*)}$ vectors, further reducing the final infidelities to $0.0005$ and $0.0003$ respectively. This may also be compared with the exact linear and harmonic motion infidelities which are $0.2773$ and $0.0062$ respectively. Crucially, the NN has learned that the MZMs have to stop before crossing the junction. The harmonic start seems more suitable for optimisation than the linear start, as it leads to slightly lower infidelity after the same amount of learning epochs and the produced motion profiles are more regular.}
    \label{fig:learning}
\end{figure}

In \Fref{fig:learning} we present the results of the numerical optimisation of the exchange motion profiles. We have first pre-trained the NN to output approximate linear motion ($\mathbf{s}_j^{(*)} = \tau$) or approximate harmonic motion ($\mathbf{s}_j^{(*)} = \sin^2(\pi \tau/2)$). It was necessary to pre-train the NN, because initialising it with random weights resulted with exchange motion profiles with quantum fidelity numerically equal to $0.0$ and its gradient exhibiting a large plateau. The NN training followed by direct gradient descent in the $\mathbf{s}_j^{(*)}$-space allowed us to reduce the quantum infidelity by several orders form $1-\mathcal{F}\sim 10^{-1}$ to $1-\mathcal{F}\sim 10^{-4}$. Crucially, the NN has learned that the MZMs have to stop before crossing the junction. This is due to the gap jump effect which we explain in Section \ref{sec:gap}. The system size is  $3N+1$ with $N=55$ and $\Delta_h \equiv \Delta =0.55$, $\Delta_v = i\Delta$, $w=2.0$, $\mu_0=1.0$, $V_0 = 30.1$, and $x_0=5$. The evolution time for each stage is $T_{stage} = 250$ and consists of $2000$ timesteps. This set of parameters puts us well into the super-adiabatic regime as $T_{stage}> 2\pi/E_{gap}\approx 15.7$ and the velocity $v = 55/250 = 0.22$ which is lower than the critical velocity $v_{crit}=\Delta$.

\subsection{Exchange in nanowires with disorder}
In realistic setups, there are different types of noise that may potentially affect the above results. In particular, the presence of disorder in the nanowire will make the base potential $\mu_0$ noisy \cite{microsoft,PhysRevLett.130.116202,dassarma_RNN}. To model this, we add a Gaussian noise term to the base potential that makes $\mu_0$ vary slightly from site to site
\[\mu_{0,j} = \mu_0 + \nu_j,\quad \nu_j\sim\mathcal{N}(0,\sigma_\nu^2).\]
We choose the same set parameters as in the previous section ($\mu_0=1.0)$ and set the noise variance to $\sigma_\nu = 0.02$. We have retrained the NN models following two different scenarios. In the first scenario, we assume that we have access to the exact noise pattern which is fixed throughout the entire training process. Experimentally, this means assuming that we are able to precisely measure the disorder pattern in the given nanowire sample. Understanding the disorder in hybrid superconductor-semiconductor nanowires has been recognised as one of the key challenges in the realisation of MZMs in solid state platforms \cite{microsoft,dassarma_RNN}. There are theoretical proposals showing that this may be accomplished by using the tunnel conductance data processed by machine learning techniques \cite{dassarma_RNN}. In the second scenario, we assume no knowledge about the disorder, so an appropriate way of optimisation in this case is to change the noise pattern after each NN training epoch/gradient step. In machine learning this is known as the online stochastic gradient descent method \cite{shalev-shwartz_ben-david_2014} applied to a sample of systems with different disorder patterns. The results of training in the above two scenatios are summarised in \Tref{tab:training_results}. The training always consists of $120$ episodes of NN training with the learning rate $10^{-4}$ and $60$ direct gradient descent steps in the $\mathbf{s}_j^{(*)}$-space with the learning rate $10^{-6}$. The results show that linear and harmonic shuttle protocols can be significantly optimised whenever the sample disorder is accurately known. However, in the case when the disorder pattern is not known, our optimisation strategy (at least with the applied number of training epochs and the learning rates) does not outperform the simple harmonic shuttle protocol. This shows that knowing the disorder pattern in the nanowire sample allows one to improve the quantum gate generation fidelity by two orders of magnitude.
\begin{table}[!htbp]
\centering
\begin{tabular}{c|c|ccc}
\toprule
\multirow{2}{*}{Start} &  No training & \multicolumn{3}{c}{Training}\\
   &  No noise & No noise    & Constant noise   & Variable noise\\
   \midrule
Linear   & $0.277$  &  $5\cdotp 10^{-4}$  & $6\cdotp 10^{-4}$  & $0.008$  \\
Harmonic   &  $0.018$ & $3\cdotp 10^{-4}$  & $9\cdotp 10^{-4}$  & $0.023$ \\
\bottomrule
\end{tabular}
\caption{The infidelities resulting from MZM transport profile optimisation for different types of noise and different starting shuttle protocols. The training consists of $120$ episodes of NN training with the learning rate $10^{-4}$ and $60$ direct gradient descent steps in the $\mathbf{s}_j^{(*)}$-space with the learning rate $10^{-6}$. The system parameters are the same as specified in \Fref{fig:learning}. The variance of the disorder noise is $\sigma_\nu = 0.02$. The learning protocol adapts efficiently to a known constant noise pattern, but fails to outperform the simple harmonic shuttle protocol (with stops at the junction) when the noise is unknown, i.e. allowed to vary during learning.}
\label{tab:training_results}
\end{table}

Furthermore, we have compared the average performance of the different trained models on a sample of $30$ disorder patterns drawn from the Gaussian distribution with the variance $\sigma_\nu = 0.02$. The results presented in \Tref{tab:sampling_results} confirm our previous conclusions that understanding the disorder pattern in the nanowire is necessary for accomplishing high fidelity quantum gate generation. We also conclude that the harmonic shuttle protocol with stops at the junction is a reasonable choice of MZM exchange protocol for systems with unknown disorder and a suitable starting point for optimisation when the disorder pattern is known.
\begin{table}[!htbp]
\centering
\begin{tabular}{c|c|ccc}
\toprule
\multirow{2}{*}{Start} &  \multirow{2}{*}{No training}  & \multicolumn{3}{c}{Training}\\
   &   & No noise    & Constant noise   & Variable noise\\
   \midrule
Linear   & $0.265\pm 0.016$  &  $0.010\pm 0.003$  & $0.015\pm 0.004$  & $0.010\pm 0.002$ \\
Harmonic   &  $0.016 \pm 0.006$ & $0.027\pm 0.007$  & $0.033\pm 0.007$   & $0.022\pm 0.005$  \\
\bottomrule
\end{tabular}
\caption{The infidelities for the different trained models averaged over a sample of $30$ disorder patterns drawn from the Gaussian distribution with the variance $\sigma_\nu = 0.02$. Other system parameters are the same as specified in \Fref{fig:learning}. The training consists of $120$ episodes of NN training with the learning rate $10^{-4}$ and $60$ direct gradient descent steps in the $\mathbf{s}_j^{(*)}$-space with the learning rate $10^{-6}$. The results show that when the disorder is unknown, the simple harmonic shuttle protocol with stops at the junction or models trained from the linear motion start may be a suitable choice. In the case of the harmonic start, we can also clearly see the effects of overfitting when the model is fitted to a particular disorder pattern, but applied to a sample of systems with varying disorder.}
\label{tab:sampling_results}
\end{table}

\section{The jump of the energy gap near the junction point}\label{sec:gap}
\begin{figure}
    \centering
    \includegraphics[width=0.85\textwidth]{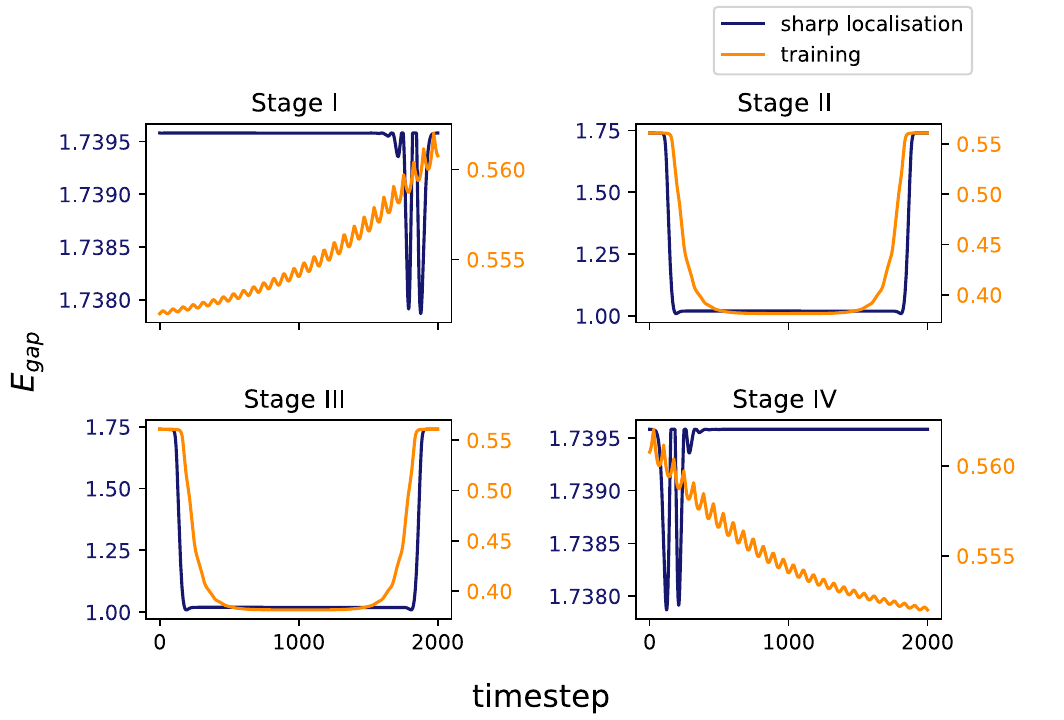}
    \caption{The behaviour of the energy gap for the system of the size $3N+1$ with $N=60$ for two different sets of parameters. The blue line labelled as {\textit{sharp localisation}} corresponds to $\Delta=1.8$, $w=2.0$, $\mu_0=0.0$, $V_0 = 30.1$, $x_0=5$ where the MZMs are sharply localised. The orange line labelled as {\textit{training}} corresponds to $\Delta=0.55$, $w=2.0$, $\mu_0=1.0$, $V_0 = 30.1$, $x_0=5$ which is the same set of parameters that was used for NN training in Section \ref{sec:learning}. One can see different types of energy gap oscillations, depending on how sharply the MZMs are localised.}
    \label{fig:gaps}
\end{figure}

As we have pointed out in Section \ref{sec:learning} and \Fref{fig:learning}, in the optimised exchange protocols the MZMs are stopping at the junction. This is due to the sharp drop of the energy gap which starts when one of the MZMs overlaps with the junction point (site $N+1$ in our labelling convention). Qualitatively, the energy gap behaves in a complicated way when one of the MZMs approaches the junction point. In particular, the gap starts to oscillate when the transported MZM approaches the junction and then drops sharply when the MZM passes the junction (see \Fref{fig:gaps}). Consequently, the time derivative of the system's Hamiltonian becomes large in this situation and the only way to mitigate this and maintain the approximate adibaticity of the time evolution is for the MZMs to slow down and effectively stop before crossing the junction point. In this section, we explain the drop in the energy gap  in the completely localised regime, i.e. $|\Delta|=w$ and $\mu = 0$. The oscillations seem to be more difficult to explain analytically as they are present only in the settings where the MZMs have some nonzero localisation length.

To explain the energy gap, we consider two $p$-wave chains of equal lengths (chain $L$ and chain $R$) that are initially decoupled. Both chains have the parameters $w = |\Delta|$ and $\mu=0$. Their hamiltonians read
\begin{equation}\label{eq:ham_LR}
\eqalign{
& H_{X}  = - \Delta  \sum_{j=1}^{N-1} \left(c_{X,j}^\dagger c_{X,j+1} + c_{X,j+1}^\dagger c_{X,j}\right) \cr 
& + \Delta \sum_{j=1}^{N-1} \left(e^{i\phi_X}\, c_{X,j} c_{X,j+1} + e^{-i\phi_X}\, c_{X,j+1}^\dagger c_{X,j}^\dagger\right), \quad \Delta>0.
 }
\end{equation}
where $X=L,R$. To simplify the calculations, we will assume that $\phi_L = 0$ and $\phi_R \equiv \phi\in[0,2\pi[$. Recall that the $L/R$ Hamiltonians can be diagonalised in the Majorana representation \cite{Kitaev_chain}
\begin{equation}
\fl \gamma_{2j-1}^{(X)} = e^{-i\phi_X/2} c_{X,j}^\dagger+^{i\phi_X/2} c_{X,j},\quad \gamma_{2j}^{(X)} = i\left(e^{-i\phi_X/2} c_{X,j}^\dagger-e^{i\phi_X/2} c_{X,j}\right).
\end{equation}
Then, we have
\begin{equation}
H_X = i\Delta \sum_{j=1}^{N-1} \gamma_{2j}^{(X)}\gamma_{2j+1}^{(X)}.
\end{equation}
In particular, the four MZMs $\gamma_{1}^{(X)}$, $\gamma_{N}^{(X)}$, $X=L,R$ are the edge modes that do not enter $H_X$. The energy gap of this system is equal to 
\begin{equation}
E_{gap,uncoupled} = \Delta.
\end{equation}

Let us next couple site $N$ of chain $L$ with site $1$ of chain $R$ using the coupling term \eref{eq:coupling}, i.e.
\begin{equation}
\fl H_{L-R} = - \Delta \left(c_{L,N}^\dagger c_{R,1} + c_{R,1}^\dagger c_{L,N}\right) + \Delta \left(e^{i\phi}\, c_{L,N} c_{R,1} + e^{-i\phi}\, c_{R,1}^\dagger c_{L,N}^\dagger\right),
\end{equation}
so that the Hamiltonian of the entire system reads $H = H_L+H_R+H_{L-R}$. The coupling Hamiltonian in terms of the Majorana operators reads
\begin{equation}
H_{L-R} = i\Delta\left(\sin\left(\frac{\phi}{2}\right)\, \gamma_{2N-1}^{(L)}\gamma_{1}^{(R)}+\cos\left(\frac{\phi}{2}\right)\, \gamma_{2N}^{(L)}\gamma_{1}^{(R)}\right).
\end{equation}
Thus, in order to diagonalise the entire system, it is enough to diagonalise the part
\begin{equation}
\fl
\eqalign{
  H_{eff} = H_{L-R}+i\Delta\,\gamma_{2N-2}^{(L)}\gamma_{2N-1}^{(L)} = \frac{i}{2}\Delta \times \cr
\pmatrix{
\gamma_{2N-2}^{(L)} & \gamma_{2N-1}^{(L)} & \gamma_{2N}^{(L)} & \gamma_{1}^{(R)}
}
\pmatrix{
0 & 1 & 0 & 0 \cr
-1 & 0 & 0 & \sin\frac{\phi}{2} \cr
0 & 0 & 0 & \cos\frac{\phi}{2} \cr
0 & -\sin\frac{\phi}{2} & - \cos\frac{\phi}{2} & 0
}
\pmatrix{
\gamma_{2N-2}^{(L)} \cr
\gamma_{2N-1}^{(L)} \cr
\gamma_{2N}^{(L)} \cr
\gamma_{1}^{(R)}
}.
}
\end{equation}
The above matrix can be diagonalised analytically and the resulting eigenenergies read $\pm\Delta\sqrt{1\pm\sin\frac{\phi}{2}}$. Thus, we have
\begin{equation}\label{eq:gap_jump}
\frac{E_{gap,coupled}}{E_{gap,uncoupled}} = \sqrt{1-\sin\frac{\phi}{2}}\leq1.
\end{equation}
In particular, when $\phi = \pi$ the two chains form the so-called $\pi$-junction where the gap closes and the two MZMs remain localised at the junction points despite the presence of the coupling. This remains true also outside the completely localised regime  \cite{Alicea2011}. However, when $\phi \neq \pi$, then the Majoranas $\gamma_{2N}^{(L)}$ and $\gamma_{1}^{(R)}$ no longer have zero energy and the entire system has just two MZMs localised at the endpoints of the connected $L-R$ chain. As we can see in \Fref{fig:gaps}, the qualitative features of the energy gap jump remain true even outside the completely localised regime where the \Eref{eq:gap_jump} provides a rough estimate for the amplitude of the gap jump. There, we have $\phi = \pi/2$, so if the MZMs were perfectly localised, the gap in the middle of stage $II$ of the exchange would be equal to $\sqrt{1-1/\sqrt{2}}\approx 0.54$ times the gap in the middle of stage $I$.

\section{Discussion and conclusions}

In this work, we have studied the problem of optimising the finite-time control protocols for Majorana zero mode exchange. To this end, we have derived an analytic formula for the gradient of quantum fidelity which allowed us to build a scalable deep learning system for the control protocol optimisation. We have worked in the super-adiabatic regime and focused on the exchange of two MZMs on a trijunction consisting of $3N+1$ sites. We have observed that the optimised exchange protocols were characterised  by stopping of the MZMs a the junction point. We have explained this stopping effect by the behaviour of the energy gap which exhibits a sharp jump when one of the MZMs approaches the junction. Our optimised protocols improve the fidelity of quantum gate generation by two orders of magnitude when compared with the simple harmonic motion shuttle protocol. However, adding unknown disorder to the nanowire causes our protocols to lose their robustness due to the overfitting. This might be remedied to some extent by applying the learning via online stochastic gradient descent for a larger number of learning epochs, possibly with decaying learning rate. This shows that understanding the disorder pattern in a nanowire is necessary for accomplishing high fidelity quantum gate generation and passing the error correction thresholds.

A natural direction of generalising our results would be to consider the exchange of two MZMs in a system consisting of the total of four MZMs (two separate topological regions). This would allow us to directly simulate a topological qubit. However, this would require further optimisation of our code as well as having access to more powerful computational resources. This is because calculating the gradient of quantum fidelity, even with the analytic formula at hand, still requires significant computational resources. For the trijunction consisting of $166$ sites ($N=55$) and time evolution of $8000$ time steps, evaluating the gradient with our current implementation required around $100$ Gigabytes of RAM and took about $45$ minutes when using $28$ CPU cores of an HPC node (so, $120$ epochs of NN training takes about four days). Realising a similar calculation for a system of four MZMs which would consist of $200-300$ sites would take a few times more resources since calculating the gradient requires several steps (such as matrix diagonalisation) which scale polynomially with the system size.

Nevertheless, our presented results do apply to systems with more than two MZMs whenever only two MZMs localised at the edges of the same topological region are exchanged and the remaining MZMs are sufficiently separated. Such exchanges are also crucial elements of quantum gate generation algorithms \cite{TQC_review,Georgiev06}.

Another possible extension of our work would concern the proximity coupled nanowires with induced $s$-wave superconductivity 
\cite{PhysRevB.88.155420,PhysRevB.85.140513}. On the technical level, this would also require more computational resources, since including spin makes the Hamiltonian twice as large. Since $p$-wave superconductors are a limiting case of $s$-wave superconductors, we anticipate similar effects concerning the stopping of MZMs at the junction point due to the presence of an analogous energy gap jump. 

\ack
The authors would like to thank Luuk Coopmans for helpful discussions, especially for suggesting us to analytically calculate the quantum infidelity gradient. We also thank Domenico Pellegrino for useful discussions.

\section*{References}
\bibliography{iopart-num}

\providecommand{\newblock}{}
\begin{thebibliography}{10}
\expandafter\ifx\csname url\endcsname\relax
  \def\url#1{{\tt #1}}\fi
\expandafter\ifx\csname urlprefix\endcsname\relax\def\urlprefix{URL }\fi
\providecommand{\eprint}[2][]{\url{#2}}

\bibitem{TQC_review}
Nayak C, Simon S~H, Stern A, Freedman M and Das~Sarma S 2008 {\em Rev. Mod.
  Phys.\/} {\bf 80}(3) 1083--1159

\bibitem{topo_book}
Simon S~H 2023 {\em Topological Quantum\/} (Oxford, UK: Oxford University
  Press) ISBN 9780198886723

\bibitem{PhysRevLett.98.010506}
Tewari S, Das~Sarma S, Nayak C, Zhang C and Zoller P 2007 {\em Phys. Rev.
  Lett.\/} {\bf 98}(1) 010506

\bibitem{PhysRevLett.99.037001}
Tewari S, Das~Sarma S and Lee D~H 2007 {\em Phys. Rev. Lett.\/} {\bf 99}(3)
  037001

\bibitem{PhysRevB.76.104516}
Grosfeld E, Cooper N~R, Stern A and Ilan R 2007 {\em Phys. Rev. B\/} {\bf
  76}(10) 104516

\bibitem{doi:10.1126/science.aan4596}
Zhang P, Yaji K, Hashimoto T, Ota Y, Kondo T, Okazaki K, Wang Z, Wen J, Gu G~D,
  Ding H and Shin S 2018 {\em Science\/} {\bf 360} 182--186

\bibitem{doi:10.1126/science.aao1797}
Wang D, Kong L, Fan P, Chen H, Zhu S, Liu W, Cao L, Sun Y, Du S, Schneeloch J,
  Zhong R, Gu G, Fu L, Ding H and Gao H~J 2018 {\em Science\/} {\bf 362}
  333--335

\bibitem{PhysRevB.92.115119}
Wang Z, Zhang P, Xu G, Zeng L~K, Miao H, Xu X, Qian T, Weng H, Richard P,
  Fedorov A~V, Ding H, Dai X and Fang Z 2015 {\em Phys. Rev. B\/} {\bf 92}(11)
  115119

\bibitem{PhysRevB.93.115129}
Wu X, Qin S, Liang Y, Fan H and Hu J 2016 {\em Phys. Rev. B\/} {\bf 93}(11)
  115129

\bibitem{PhysRevLett.117.047001}
Xu G, Lian B, Tang P, Qi X~L and Zhang S~C 2016 {\em Phys. Rev. Lett.\/} {\bf
  117}(4) 047001

\bibitem{Kong2021}
Kong L, Cao L, Zhu S, Papaj M, Dai G, Li G, Fan P, Liu W, Yang F, Wang X, Du S,
  Jin C, Fu L, Gao H~J and Ding H 2021 {\em Nature Communications\/} {\bf 12}
  4146 ISSN 2041-1723

\bibitem{PhysRevLett.105.077001}
Lutchyn R~M, Sau J~D and Das~Sarma S 2010 {\em Phys. Rev. Lett.\/} {\bf 105}(7)
  077001

\bibitem{PhysRevLett.105.177002}
Oreg Y, Refael G and von Oppen F 2010 {\em Phys. Rev. Lett.\/} {\bf 105}(17)
  177002

\bibitem{Alicea2011}
Alicea J, Oreg Y, Refael G, von Oppen F and Fisher M~P~A 2011 {\em Nature
  Physics\/} {\bf 7} 412--417 ISSN 1745-2481

\bibitem{Das2012}
Das A, Ronen Y, Most Y, Oreg Y, Heiblum M and Shtrikman H 2012 {\em Nature
  Physics\/} {\bf 8} 887--895 ISSN 1745-2481

\bibitem{doi:10.1126/science.1222360}
Mourik V, Zuo K, Frolov S~M, Plissard S~R, Bakkers E~P~A~M and Kouwenhoven L~P
  2012 {\em Science\/} {\bf 336} 1003--1007

\bibitem{doi:10.1126/science.1259327}
Nadj-Perge S, Drozdov I~K, Li J, Chen H, Jeon S, Seo J, MacDonald A~H, Bernevig
  B~A and Yazdani A 2014 {\em Science\/} {\bf 346} 602--607

\bibitem{Xu2016}
Xu J~S, Sun K, Han Y~J, Li C~F, Pachos J~K and Guo G~C 2016 {\em Nature
  Communications\/} {\bf 7} 13194 ISSN 2041-1723

\bibitem{Kitaev_chain}
Kitaev A~Y 2001 {\em Physics-Uspekhi\/} {\bf 44} 131

\bibitem{PhysRevB.84.035120}
Clarke D~J, Sau J~D and Tewari S 2011 {\em Phys. Rev. B\/} {\bf 84}(3) 035120

\bibitem{PhysRevA.71.022316}
Bravyi S and Kitaev A 2005 {\em Phys. Rev. A\/} {\bf 71}(2) 022316

\bibitem{PhysRevB.88.064515}
Scheurer M~S and Shnirman A 2013 {\em Phys. Rev. B\/} {\bf 88}(6) 064515

\bibitem{PhysRevB.91.201404}
Karzig T, Rahmani A, von Oppen F and Refael G 2015 {\em Phys. Rev. B\/} {\bf
  91}(20) 201404

\bibitem{PhysRevB.100.134307}
Conlon M, Pellegrino D, Slingerland J~K, Dooley S and Kells G 2019 {\em Phys.
  Rev. B\/} {\bf 100}(13) 134307

\bibitem{LuukNN}
Coopmans L, Luo D, Kells G, Clark B~K and Carrasquilla J 2021 {\em PRX
  Quantum\/} {\bf 2}(2) 020332

\bibitem{github_repo}
Maciazek T and Conlon M 2023 Optimising the exchange of majorana zero modes in
  a quantum nanowire network, {G}it{H}ub repository
  \urlprefix\url{https://github.com/tmaciazek/trijunction\_mzm\_braiding}

\bibitem{Schmidt2019}
Schmidt J, Marques M~R~G, Botti S and Marques M~A~L 2019 {\em npj Computational
  Materials\/} {\bf 5} 83 ISSN 2057-3960

\bibitem{Bedolla_2021}
Bedolla E, Padierna L~C and Castañeda-Priego R 2020 {\em Journal of Physics:
  Condensed Matter\/} {\bf 33} 053001

\bibitem{dawid2022modern}
Dawid A {\em et~al.\/} 2022 Modern applications of machine learning in quantum
  sciences (\textit{Preprint} \eprint{2204.04198})

\bibitem{PhysRevLett.130.116202}
Thamm M and Rosenow B 2023 {\em Phys. Rev. Lett.\/} {\bf 130}(11) 116202

\bibitem{dassarma_RNN}
Taylor J~R, Sau J~D and Das~Sarma S 2023  (\textit{Preprint}
  \eprint{arXiv:2307.11068})

\bibitem{PhysRevLett.125.170501}
Zhang Y~H, Zheng P~L, Zhang Y and Deng D~L 2020 {\em Phys. Rev. Lett.\/} {\bf
  125}(17) 170501

\bibitem{10.21468/SciPostPhys.5.1.004}
Bauer B, Karzig T, Mishmash R~V, Antipov A~E and Alicea J 2018 {\em SciPost
  Phys.\/} {\bf 5} 004

\bibitem{PhysRevB.107.104516}
Truong B~P, Agarwal K and Pereg-Barnea T 2023 {\em Phys. Rev. B\/} {\bf
  107}(10) 104516

\bibitem{BdGbook}
Ring P and Schuck P 1980 {\em The Nuclear Many-Body Problem\/} (Heidelberg:
  Springer Berlin) p 718

\bibitem{Onishi66}
Onishi N and Yoshida S 1966 {\em Nuclear Physics\/} {\bf 80} 367--376 ISSN
  0029-5582

\bibitem{Beck1970}
Beck R, Mang H~J and Ring P 1970 {\em Zeitschrift f{\"u}r Physik A Hadrons and
  nuclei\/} {\bf 231} 26--47 ISSN 0939-7922

\bibitem{JB76}
Jennrich R~I and Bright P~B 1976 {\em Technometrics\/} {\bf 18} 385--392 ISSN
  00401706

\bibitem{KL85}
Kalbfleisch J~D and Lawless J~F 1985 {\em Journal of the American Statistical
  Association\/} {\bf 80} 863--871 ISSN 01621459

\bibitem{TC03}
Tsai H and Chan K~S 2003 {\em Bernoulli\/} {\bf 9} 895--919 ISSN 13507265

\bibitem{Adam}
Kingma D~P and Ba J 2015 Adam: A method for stochastic optimization {\em ICLR
  2015 - Conference Track Proceedings\/} (\textit{Preprint}
  \eprint{arXiv:1412.6980})

\bibitem{microsoft}
Aghaee M {\em et~al.\/} (Microsoft Quantum) 2023 {\em Phys. Rev. B\/} {\bf
  107}(24) 245423

\bibitem{shalev-shwartz_ben-david_2014}
Shalev-Shwartz S and Ben-David S 2014 {\em Understanding Machine Learning: From
  Theory to Algorithms\/} (Cambridge University Press)

\bibitem{Georgiev06}
Georgiev L~S 2006 {\em Phys. Rev. B\/} {\bf 74}(23) 235112

\bibitem{PhysRevB.88.155420}
Pientka F, Glazman L~I and von Oppen F 2013 {\em Phys. Rev. B\/} {\bf 88}(15)
  155420

\bibitem{PhysRevB.85.140513}
Lutchyn R~M, Stanescu T~D and Das~Sarma S 2012 {\em Phys. Rev. B\/} {\bf
  85}(14) 140513

\end{thebibliography}

\end{document}